\documentstyle[12pt,epsf]{article}
\def \aa    #1 #2   {{ A\&A \/}, {#1}, {#2}}
\def \aas   #1 #2   {{ A\&AS \/}, {#1}, {#2}}
\def \aj    #1 #2   {{ AJ \/}, {#1}, {#2}}
\def \apj   #1 #2   {{ ApJ \/}, {#1}, {#2}}
\def \apjs  #1 #2   {{ ApJS \/}, {#1}, {#2}}
\def \mnras #1 #2   {{ MNRAS \/}, {#1}, {#2}}
\def \prl   #1 #2   {{ Phys. Rev. Lett. \/}, {#1}, {#2}}
\def \nat   #1 #2   {{ Nature \/}, {#1}, {#2}}
\def \com   #1 #2   {{ Comments Astrophys.\/}, {#1}, {#2}}
\def \sast  #1 #2   {{ Soviet Astron. \/} {#1}, {#2}}
\def \sastl #1 #2   {{ Soviet Astron. Lett.\/}, {#1}, {#2}}
\def \astr  #1 #2   {{ Astrophysics \/}, {#1}, {#2}}
\begin{document}
\title{TOPOLOGY OF THE GALAXY DISTRIBUTION \footnote{ to be published in
Proc. XV-th Moriond Astrophysics Meeting on {\it Clustering in the Universe}}}

\author{Sergei F. Shandarin}
       { University of Kansas, Lawrence, U.S.A.}

\begin{abstract}{\baselineskip 0.4cm}

The history and the major results of the study of the topology
of the large-scale structure are briefly reviewed. Two techniques based
on percolation theory and the genus curve are discussed.
The preliminary results
of the percolation analysis of the Wiener reconstruction of the
IRAS $1.2 Jy$ redshift catalog are reported.
\end{abstract}

\section{Introduction}
Looking at the large-scale distributions of galaxies, one may notice
that along with clusters and groups of galaxies there are also
conspicuously oblong concentrations of galaxies: filaments. One may also
get an impression that filaments are connected to each other forming a
single network spanning through the entire sample.
The two-point
correlation function (the most common way of describing the distribution
of galaxies) obviously is not sensitive to the geometry and topology of the
galaxy distribution.
The three- and many- point correlation functions generally speaking
are sensitive to the shapes and probably the topology
but become cumbersome very quickly.
Currently popular averaged moments are easy to interpret, but they loose
sensitivity to the geometry after averaging over the volume \cite{Bern95}.
There have been suggested various statistics to characterize the geometry
and topology of the large-scale structure.
In this talk I briefly review the studies of the topology
of the large-scale galaxy distribution.

The first mention of topology in the context of the large-scale
structure problem (I am familiar with) was in the 1970 paper by Doroshkevich
\cite{Dor70}.
Studying the formation of the large-scale structure in the pancake
scenario Doroshkevich calculated the Euler characteristic
of the isodensity surfaces of the initial
{\it Gaussian} density field.
Both the Euler characteristic and a common topological measure used
in cosmology, genus, are
determined by the mean Gaussian curvature of the surface of a
constant density.

In 1982 Zel'dovich noticed that the percolation properties of the
{\it nonlinear} density distribution in the HDM
(Hot Dark Matter) model are very different from the initial Gaussian field.
He also suggested characterizing the topology of the nonlinear density
distribution by the percolation thresholds \cite{Z82}.
Percolation theory deals with the number and properties of the ``clusters'',
which are defined as connected regions bounded by the surfaces of a
constant density. Following Zel'dovich's
idea the author of this talk suggested to use percolation properties
of {\it the galaxy distribution} as an objective quantitative measure of the
topology of the large-scale structure and also as a discriminator
between cosmological models \cite{Sh83}, \cite{Sh-Z83}.

The percolation technique was utilized in the study of the CfA I catalog
\cite{E-K-S-Sh84}. It was found that the large-scale
distribution
of galaxies had a network structure. Theoretical studies of the models
with the power law initial spectra showed that the $n=-1$ model clearly
percolated better than the $n=0$ model and also in the  $\Omega=1$ universe
the $n=-1$ model was in an agreement with the observations \cite{B-B83}.
The percolation
method showed that the CDM (Cold Dark Matter) model appeared filamentary
rather than hierarchical \cite{M-etal83}, \cite{D-etal85}.
It was also pointed out that the major disadvantage of percolation
technique was the dependence
of the percolation thresholds on the mean density of the sample \cite{D-W85}
which made it difficult to apply to sparse samples. Similarly, we note that at
present some believe that sparse samples can be reliably used for
the estimation of the two-point correlation function only \cite{Bouch95}.

In 1986 Bardeen et al \cite{Bard-etal86} and
Gott, Melott and Dickinson rediscovered Doroshkevich's
idea of utilizing the Euler characteristic and expanded it
to the {\it nonlinear} distributions
as well as {\it galaxy} catalogs \cite{G-M-D86}.
(Both percolation and genus techniques assumes some kind of smoothing when
applied to galaxy distributions.)
However, instead of
the mean density of the Euler characteristic $\chi$ used by Doroshkevich
they introduced  the mean density of genus, $g$, which is
proportional to the Euler characteristic: $g=-\chi/2$.
(For a general review of this method see e.g. \cite{Mel90}.)
Tomita \cite{Tom86} gave a very elegant analytic expression for the mean Euler
characteristic for a D-dimensional Gaussian field which in three-dimensional
space yields the familiar equation for the mean genus density

\begin{equation}
g = {1 \over 4\pi^2}\Bigl({<k^2> \over 3}\Bigr)^{3/2}(1-\nu^2)\exp(-\nu^2/2),
\label{eq:genus} \end{equation}

where $\nu=\delta/\sigma_{\delta}$ is the number of standard
deviations by which the threshold density
departs from the mean density,
$\delta \equiv {(\rho -\bar{\rho})}/\bar{\rho}$,
$<k^2> = \int k^2P(k)d^3k/\int P(k)d^3k$,
and $P(k)$ is the power spectrum (see e.g. \cite{Vog-etal94}).
In a Gaussian field the genus curve has a maximum
at $\nu =0$ with the amplitude determined by
the characteristic scale of the density field $<k^2>$.
Since it depends on the
slope of the spectrum it is often used as a measure of the
``effective'' slope of
the spectrum (see e.g. \cite{Moore-etal92});
however, it is worth mentioning that $<k^2>$ has a stronger
dependence on the smoothing scale than on the spectral index.
Therefore even weak nonlinearity on the smoothing scale may influence
the estimate of the spectral index.
The genus curve changes sign two times at $\nu =\pm 1$ which signifies
the qualitative change of the topology of the surface separating
high ($\delta>\delta_c$) and low ($\delta<\delta_c$) density regions where
$\delta_c$ is a chosen density threshold.
As suggested by  Eq.\ref{eq:genus} there are only two types of
qualitatively different topologies: 1) one phase percolates and the other
does not (negative genus)
and 2) both phases percolate (positive genus).
In the range
$-1< \nu <1$ both phases percolate through the whole region and this is
often reffered to as sponge topology. At $\nu <1$ the low density regions
do not percolate and at $\nu >1$ the high density regions do not percolate.
The topology of the separating surfaces is obviously the same
at $\nu =\pm\nu_c$ as measured by Eq.\ref{eq:genus}.
But in cosmological literature it is labeled either as a meatball or bubble
topology depending on whether the high or
low density phase does not percolate. The network structure obviously
has a sponge topology however the term emphasizes a geometrical aspect
of a non-Gaussian density field: the high density regions at the percolation
threshold occupy a smaller volume than that of a Gaussian field.

The change of the genus sign is believed to coincide with the percolation
thresholds however there is no theorem proving that. Intuitively, it is
plausible for distributions resulting from Gaussian fields
due to gravitational instability
if one believes in a common interpretation of the genus
as the mean density of the number of holes minus the number of the isolated
regions and that clusters with holes inside do not form. Under these
conditions, we
assume that percolation thresholds coincide with the changes of the genus
sign at least approximately.

One advantage of the genus method is the existence of the analytic expression
for Gaussian random fields (Eq.\ref{eq:genus}). Recently there has also been
an analytic expression obtained in the weakly nonlinear regime \cite{Mats94}.
However, one should not forget that the mean genus is a statistical
measure and therefore an estimate of errors is needed before it becomes
meaningful. The errors for finite samples having finite resolution
can be estimated only from numerical simulations. Percolation parameters
are also calculated numerically, but if one can estimate the errors he
almost certainly can estimate the mean with similar accuracy.

It has been claimed that the percolation thresholds are the most sensitive
discriminators of the models \cite{Sh83}. The recent study of the
CfA II catalog using the genus method \cite{Vog-etal94}
seems to support that suggestion.
The authors reduced the
information of the genus curve to three numbers one of which was
the genus peak width $W_{\nu}=\nu_+ -\nu_-$ where $\nu_+$ and
$\nu_-$ are the levels at which the genus changes the sign.
Fig. 12 through 14 in \cite{Vog-etal94} clearly
demonstrate that $W_{\nu}$ has the highest discriminating power.
However,
we still believe that the percolation thresholds, $\nu_+$ and $\nu_-$,
should be interpreted separately because
they carry independent information about the topology of the structure.

\section{Largest ``cluster'' and largest ``void''}
Percolation theory deals with the number and properties of the ``clusters''.
In the absence of a better term we label as ``clusters'' the regions bounded
by the surfaces of chosen constant density. In order to avoid confusion
with clusters of galaxies, we will use quotation marks when talking about
``clusters'' with this non-astronomical meaning. The density threshold
$\delta_c$ separating
high ($\delta>\delta_c$) and low ($\delta<\delta_c$) density regions is
assumed to be a free parameter $\delta_c >-1$. Analyzing discrete distributions
(e.g. galaxy distributions) we assume a smoothing procedure creating
a continuous density distribution.

At every density threshold all ``clusters'' and ``voids'' are identified and
various types of analysis can be performed \cite{K-Sh93}.
However here we present only
the results of the study of the largest ``cluster'' and the largest ``void'' as
functions of a density threshold.
(The largest ``void'' is defined as the largest ``cluster'' in the low density
phase.) The full analysis will be presented elsewhere \cite{Y-Sh95},
\cite{Y-Sh-F95}.
The choice of the largest structure is determined by the fact that at the
percolation threshold the largest structure become infinite which signifies
the change of topology.

The density threshold is not a convenient parameter
if linear (Gaussian) and nonlinear density distributions are to be compared.
Instead we utilize the filling factor to parameterize the density threshold
\cite{dL-G-H91}.
The filling factor is the fraction of the volume occupied by the given phase.
In this case one can easily compare the properties of ``clusters'' with that
of ``voids'' and also linear and nonlinear density distributions. It is similar
to comparing different patterns provided that the same amount of paint was
used to make each pattern. The filling factor as a function of the density
threshold is obviously the cumulative distribution function.

The largest ``cluster'' and ``void'' are measured as a fraction of
the corresponding filling factor. Thus if the largest ``cluster'' is $0.9$ at
filling factor of $0.2$ it means that the density is higher than the chosen
threshold in $20\%$ of the volume and almost all of that volume
($90\%$) comprised of only one connected region.

The top two panels of Fig.1 show the largest ``cluster''
and the largest ``void''
for the density distributions obtained in the N-body simulation of the
power law model with $n=-1$ at two stages of evolution: $\lambda_{nl}=1/8$
and $1/4 L_{box}$. The simulations have been done with $128^3$
particles on the equivalent mesh \cite{M-Sh93} but for this analysis
the mesh has been reduced to $64^3$.
Error bars show $1\sigma$ deviations from the mean obtained in four
different realizations of the model.

The qualitative behavior of both of the largest structures is universal:
at small filling
factors the largest structure is negligible then at some filling factor
it quickly grows and becomes the only significant structure in the
corresponding phase. This is the percolation transition and also the
indication of the change of topology.
In Gaussian fields there is no statistical
difference between ``clusters'' and ``voids'' and the transition happens at
a filling factor of about $16\%$ corresponding to $\nu =\pm1$.
However, a finite size of the sample
as well as finite resolution biases the transition.
In order to avoid these effects, we obtain
the ``Gaussian'' distribution
by mixing the phases of the Fourier transform of the nonlinear density
distributions in question. This automatically includes all finite
grid effects in the reference Gaussian field keeping the Fourier amplitudes
exactly the same.
This allows for the generation of as many Gaussian realizations with identical
amplitudes as needed to estimate the dispersion.
The Gaussian largest structure is shown as a dotted
line in Fig.1
(hidden by the shade of the error bars) lying between the solid and dashed
lines.

The major feature of the nonlinear distribution is that the largest
``cluster'' percolates easier and the largest
``void'' harder than in the Gaussian case.
The significance of this conclusion for the largest cluster is at the
many-$\sigma$ level (see Fig.1). Qualitatively this remains true for
all models we have
studied ($n=1,0,-1,-2,-3$, CDM, and C+HDM \cite{K-Sh93}), but quantitatively
the transitions are different. The high density
regions form a connected network spanning through the whole region when the
filling factor is relatively small (smaller than in the Gaussian case) and
therefore this transition can be labeled as a shift
toward the network structure.
On the contrary the low density regions do not form a percolating
void (remain isolated)
even when the filling factor of the low density phase
is greater than that in the
Gaussian field. This type of transition can be labeled as a shift toward
the bubble structure.
The range of the sponge topology  is typically (but not necessarily)
increased compared
to the Gaussian case. Thus the above changes also can be labeled
as a shift to a sponge topology.
However, the major point is not how to label a structure but rather
to show that in a general case the two shifts are independent of each other
and carry independent information about the structure. Therefore combining
them into one parameter (like $W_{\nu}=\nu_+ - \nu_-$ mentioned above)
results in lost information.

At small filling factors the largest structure must be negligible in
sufficiently large samples. The actual (finite) sizes of the largest structures
can be used as an internal characteristic of the fairness of the sample.

In the past the percolation technique has been successfully applied to
volume limited samples \cite{E-K-S-Sh84}.
Analyzing the statistically homogeneous distributions is very easy and the
largest structures clearly distinguish between the models \cite{Y-Sh95}.
However, the analysis is much more difficult if the distribution
has a radial gradients, like the IRAS $1.2 Jy$ redshift catalog.

\section{The IRAS $1.2 Jy$ catalog}
We analyze the whole-sky galaxy distribution in real space
reconstructed from the redshift IRAS $1.2 Jy$ catalog  using a Wiener
filter and an expansion in spherical harmonics \cite{Lah-etal94}. The Wiener
filter effectively uses a variable window size which is about $500 km/s$ at
$2000 km/s$ and increases to $1800 km/s$ at $10000 km/s$. The resulting
smoothed galaxy distribution is not statistically homogeneous.
This is the major
challenge for applying the percolation technique.

In order to test the effect of the reconstruction we generated two density
distributions from the N-body simulation ($n=-1$, $k_{nl}=8$): one with the
Wiener filtering and the other without it. The percolation statistics of
these distributions are shown in the middle and bottom  panels of Fig.1.
The panels on the
left hand side show the largest structures in the reconstructed density field
without the Wiener filter and on the right hand side with the Wiener filter
applied. The panels in the middle row show the results for the sphere with
the radius of $30$ mesh units and the bottom panels show the results for
smaller sphere with the radius of $24$ mesh units.

The middle and bottom panels show only one realization which demonstrates
a substantial change in the topological properties due to the reconstruction
procedure. We plot here only one realization for better visual comparison with
the data.
\begin{figure}[t]
     \caption{The N-body simulations of the $n=-1$ model.
The top row show two stages of evolution $\lambda_{nl} =1/8 $ and
$1/4 L_{box}$ before reconstruction.
The largest ``cluster'' (solid line) and largest ``void''
(dashed line) is shown as a function of filling factor;
the Gaussian realization
with the identical Fourier amplitudes is shown in between (shaded by the error
bars). The middle and bottom panels are the reconstructions without the Wiener
filter (on the left hand side) and with the Wiener filter (on the right hand
side). Two different radii of the sphere has been used: $R=30$ and $24$ mesh
units.}
\end{figure}
Fig.2 shows the largest structures in the filtered galaxy distribution
assuming two different $\beta =\Omega^{0.6}/b$. This preliminary result
is in a very general agreement with the $n=-1$ model. The models with
$\beta=0.1$ and $\beta=1$ do not look much different. Unfortunately the
method of estimating the dispersion described above does not work in the
inhomogeneous case. We are working now on the improved version of estimating
the dispersion.

The largest structures in the galaxy distribution are quite large at small
filling factors: $30 - 40 \%$ of the corresponding filling factor. It may mean
that the sample is not large enogh for this kind of analysis.

\begin{figure}[t]
     \caption{The Wiener reconstruction of the IRAS $1.2 Jy$ redshift
catalog in spherical harmonics. Solid lines show the largest ``cluster'' and
dashed lines show the largest ``void'' at two values of
$\beta =\Omega^{0.6}/b$  and at two radii of the spherical region.}
\end{figure}

\section{Summary}
The quantitative topology proved to be a useful technique
for studying the large-scale structure in the universe. There have been
suggested two different methods for quantitative characterization of the
topology. One widely used method measures the mean density of genus of
a constant density surface as a function of the density threshold.
It actually measures the mean Gaussian curvature of the surface.
The other more geometrical method is based on percolation theory.
Here the number and properties of the ``clusters'' are studied.
In particular the volume of the largest ``cluster'' measured
as a function of a density threshold plays an important role.

In general the questions addressed by the two methods are similar but not
identical. In the case where the questions coincide the two approaches
use different methods of solving them. In particular, they treat the
boundaries differently and are probably affected differently by noise.

The genus curve method is more developed and has been applied to many
catalogs of galaxies (both two-dimensional and  redshift surveys).
The major results indicate that.
non-Gaussian behavior has been
detected in both types of the catalogs (see e.g. \cite{Col-etal93},
\cite{Vog-etal94}, \cite{Moore-etal92} and references therein). The
characterization of the structure is somewhat conflicting. Almost all
possible labels have been assigned to the galaxy distributions: meatball,
sponge, network, bubble topology. However, it is likely that the structure
seen in different catalogs looks different. \
The estimate of effective slope is in agreement with $n=-1$
\cite{Moore-etal92}. \

The percolation method has been tested in various theoretical models and
developed to the level when it can be applied to the galaxy catalogs.
For the first time we try to apply it to the magnitude limited sample
and find the preliminary result encouraging.

{I am grateful to Capp Yess and Karl Fisher
for allowing me to
report the preliminary results of a common unfinished work and
Francis Bernardeau for useful discussions during the meeting.
I acknowledge the AAS travel grant, NSF grant AST-9021414, and
University of Kansas GRF-94 grant.}

\vfill
\end{document}